\documentclass{article}
\usepackage{amsmath, amsthm, amssymb,graphicx}
\usepackage{psfrag}
\usepackage{subfigure}
\usepackage{epsf}

\newcommand{\beqas}{\begin{eqnarray*}}
\newcommand{\eeqas}{\end{eqnarray*}}
\newcommand{\beqa}{\begin{eqnarray}}
\newcommand{\eeqa}{\end{eqnarray}}
\newcommand{\ds}{\displaystyle}
\newcommand{\be}{\begin{equation}}
\newcommand{\ee}{\end{equation}}
\newcommand{\ovR}{\overline{R}}
\newcommand{\ovx}{\overline{x}}
\newcommand{\ovy}{\overline{y}}

\newtheorem{proposition}{proposition}[section]
\newtheorem{theorem}[proposition]{Theorem}

\begin{document}
\title{Stability analysis of a max-min fair Rate Control Protocol (RCP)
in a small buffer regime}

\author{Thomas Voice \ \ ({\tt T.Voice@statslab.cam.ac.uk}) \\
Gaurav Raina \ \ ({\tt G.Raina@statslab.cam.ac.uk})}
\date{}
\maketitle

\begin{abstract} 
In this note we analyse various stability properties of the max-min
fair Rate Control Protocol (RCP) operating with small buffers. We first
tackle the issue of stability for networks with arbitrary topologies. We
prove that the max-min fair RCP fluid model is globally stable in the
absence of propagation delays, and also derive a set of conditions for local
stability when arbitrary heterogeneous propagation delays are present.
The network delay stability result assumes that, at equilibrium, there is
only one bottleneck link along each route. Lastly, in the simpler setting
of a single link, single delay model, we investigate the impact of the loss
of local stability via a Hopf bifurcation.
\end{abstract}

\noindent {\bf keywords} rcp, max-min fairness, small buffers,
stability, bifurcation.

\section{Introduction}
The Rate Control Protocol (RCP) \cite{rcp,rcp-ac} takes a radically
different approach towards managing flow and congestion control as
compared to the existing Transmission Control Protocol (TCP)
congestion avoidance framework. The TCP framework has, imbedded in
it, an implicit mechanism for detecting congestion within the
network. Loss of a packet, caused by the overflow of a buffer, is
intended to provide the necessary feedback information. In sharp
contrast, RCP aims to achieve fast flow completion times by
communicating \emph{explicit} rate feedback between routers and
end-stations \cite{nd06}. RCP is closely related to the eXplicit
Control Protocol, XCP \cite{katabi}; another algorithm which
proposes the use of explicit rate feedback. Both the RCP and XCP
algorithms intend to converge to a max-min fair resource allocation
\cite{bala2005, rcp, katabi}; see \cite{peter, mw, radu, tan} and
references therein for a sample of the literature exploring various
issues related to different notions of fairness in a networking
context.

The performance of congestion control algorithms is often coupled
with the choice of certain parameters in routers in the network. For
example, such parameters may correspond to different choices of
Active Queue Management (AQM) schemes \cite{deb, hollot, srikant1,
srikant2}, or the size of buffers in routers \cite{buffer,
raina_wischik2005}. Researchers have begun questioning the design
rules for sizing buffers in core routers in the Internet
\cite{buffers} and some recent literature \cite{enachescu2005,
raina_wischik2005, wischik2005} suggests that buffers should indeed
be much smaller as compared to the current design principles. In
light of the separate developments made towards the understanding of
RCP \cite{bala2005,nd06,rcp,rcp-ac} and buffer sizing \cite{buffers,
enachescu2005, raina_wischik2005,wischik2005} it is natural to
investigate the two, i.e., RCP and small buffers, together. Such an
investigation is also motivated by the observation that previous
buffer sizing studies have focussed primarily on the impact of
smaller buffers with TCP. In this paper, the focus of our analysis
will be on the limiting regime of a max-min fair RCP fluid model
operating over a communication network with small buffers.

Previous control theoretic analysis of RCP has focussed on a single
bottleneck, where the queue is modelled as a saturated integrator
\cite{bala2005}. In our work, we assume that the size of the buffers
is small enough so that it is no longer possible to explicitly model
the queue. Rather, with such small buffers, at the time scale of
operation of the congestion controllers, it is the distribution of
queue size that plays the important role \cite{wischik2005}. It is
worth stating that we do not, in any way, contribute to the question
of exactly how small the size of buffers ought to be. The model we
analyse is simpler than the one analysed by \cite{bala2005} in that
we do not explicitly model the queue, but more involved in the sense
that our framework represents a heterogeneous network of arbitrary
topology. In our fluid model for RCP, apart from the model for the
queue, all other parameters are exactly the same as specified in the
original RCP algorithm \cite{rcp}.

The styles of analysis we employ have all been popular in the study
of congestion control: for example, global stability without
propagation delays \cite{srikant, voice}, local stability with
delays \cite{buffer, hollot, kunniyur1, srikant1, srikant2, srikant,
glen1}, and also an analysis of the dynamic system when stability
may not be guaranteed \cite{han2005, la, raina2005}.

We now outline the essence of our contribution. First, we prove that
the fluid model of RCP is globally stable in the absence of
propagation delays. Then, we derive a set of conditions for local
stability when arbitrary heterogeneous propagation delays are
present. The network delay stability result relies upon the
assumption that, at equilibrium, there is only one bottleneck link
along each route. Furthermore, in the simpler setting of a single
link, single delay, model we are able to provide a more in depth
study of the RCP algorithm. Using bifurcation analysis, we
investigate the impact of the loss of local stability in a special
case where we show that the RCP algorithm would always give rise to
a super-critical Hopf bifurcation.

An \emph{overview} of this paper follows. In Section $2$, we analyse
the stability properties of RCP over networks of arbitrary topology.
In Section $3$, we investigate local instability of RCP in a single
link, single delay, model via a Hopf bifurcation analysis. Finally,
in Section $4$ we summarise our contribution, and discuss some
avenues for further research.

\section{RCP over arbitrary topology networks}

In this section, our objective is to analyse the stability of RCP in
a small buffer regime, over networks with an arbitrary topology. We
first show global stability for the RCP fluid model in the absence
of propagation delays and then derive a set of conditions for local
stability when arbitrary heterogeneous propagation delays are
present. The network delay stability result relies upon the
assumption that, at equilibrium, there is only one bottleneck link
along each route. That is, at equilibrium, for each route there is a
unique link with minimal per-flow bandwidth available. We begin with
the model description.

\subsection{Model description}

At the level of theoretical abstraction, our set up for an Internet
like communication network follows the commonly adopted framework
\cite{srikant}.

We suppose that the communication network comprises of an
interconnection of a set of \emph{routes}, $S$, with a set of
\emph{links}, $J$. Each route $r \in S$ represents a user of the
network. Associated with each route is a set of links which
represents the path along which that user transmits information
through the network. Further, a route $r$ has associated with it a
flow rate $x_r(t) \geq 0$, which represents a dynamic fluid
approximation to the rate at which the user is sending packets along
route $r$, at time $t$.

The flow rate, for each $r \in S$ is determined by the links $l \in
r$, via the use of \emph{explicit} rate feedback. Each link $l \in
J$ has associated with it a flow rate $R_l(t)$, which represents the
maximum flow rate allowed for routes which pass through $l$. Each
packet which is sent through the network carries, in its header, an
explicit rate feedback variable. This variable is initially set to
the maximum desired flow rate for $r$. As the packet passes through
each $l \in r$, if the feedback variable is greater than $R_l(t)$,
then link $l$ sets it equal to $R_l(t)$. When the packet reaches its
destination, an acknowledgement packet ({\tt ack}), containing the
final value of the explicit rate feedback variable, is returned to
the origin of $r$, and the flow rate $x_r(t)$ is updated
accordingly.

For each route $r$ and link $l \in r$, we let $\tau_{rl}$ denote the
propagation delay from the origin of $r$ to $l$, i.e. the length of
time it takes for a packet to travel from the origin to link $l$
along route $r$. Let $\tau_{lr}$ denote the propagation delay from
$l$ to the origin of $r$, i.e. the length of time it takes for the
explicit rate feedback information from link $l$ to reach the user
who is transmitting along route $r$. In RCP, a packet must reach its
destination before an acknowledgement packet ({\tt ack}) is returned
to its source. Furthermore, as we are interested in a network with
small buffers, we may safely assume that queuing delays are a
negligible component of the end to end delay. Thus for all $l \in
r$, $\tau_{rl} + \tau_{lr} = \tau_r$, the round trip time for route
$r$.

We now have the following model for the end-system behaviour of RCP.
For each route $r \in S$,
\be x_r(t) = \text{min}_{l \in r} R_l(t - \tau_{lr}).
\label{eqn:RCPx} \ee
Now, for each link $l \in J$, $R_l(t)$ is updated depending upon the
total aggregate flow through link $l$ at time $t$. Although the
update rule is discrete, we can model it via a fluid approximation
with the following differential equation \cite{bala2005}. For each
link $l \in J$,
\be \dot{R}_l(t) = R_l(t) \left( \frac{\alpha_l}{d_l C_l} ( C_l -
y_l(t) ) - \frac{\beta_l q_l(t)}{d_l^2 C_l} \right)^+_{R_l(t)},
\label{eqn:RCPr} \ee
where $\alpha_l, \beta_l$ are positive constants, $C_l$ is the
capacity at link $l$, $d_l$ is the average round trip time of the
\emph{flows} passing through link $l$,
\beqa y_l(t) = \sum_{r : l \in r} x_r(t - \tau_{rl}),
\label{eqn:flow} \eeqa
and $q_l(t)$ is the queue size at time $t$. Here we use the notation
$a = (b)_c^+$ to mean that $a = 0$ if $b < 0$ and $c \leq 0$,
otherwise $a = b$. In the original RCP model \cite{rcp}, $\alpha_l =
\alpha$ and $\beta_l = \beta$ for each $l \in J$ for some $\alpha$
and $\beta$. We have allowed these constants to vary between links
in order to study their effect on stability.
\\

\noindent {\bf Model for the queue.} The basis of our investigation
is to study a regime where the buffers are so small that it is no
longer possible to explicitly model the queue as a saturated
integrator. This assumption simply expresses the idea that with
small enough buffers, we may use the approximation
\beqa q_l(t) = p_l(y_l(t)), \label{eqn:smallQ} \eeqa
where $p_l(\cdot)$ is a continuously differentiable function
representing the \emph{mean queue length} of link $l$. This is
consistent with the observation in \cite{wischik2005} that in a
small buffer regime, it is the distribution of the queue size that
plays the prominent role in the dynamics of the congestion control
framework. At the level of interest in this paper, we do not
motivate any explicit functional form for the mean queue length.
Different functional forms may be suitable candidates, but our
primary focus is to investigate some stability properties of the
dynamical system defined by (\ref{eqn:RCPx}-\ref{eqn:smallQ}).

\subsection{Global stability without propagation delays}

In this subsection, we seek to investigate the stability properties
of RCP without taking into account the effects of propagation
delays. This gives us some insight into the stability of the
algorithm in general, and is also a plausible model for the special
case when propagation delays are small in comparison to the update
step size of the algorithm. In terms of the fluid model, this would
mean
$$
\max_{l \in J} \frac{\alpha_l}{d_l C_l} + \frac{\beta_l
q_l(t)}{d_l^2 C_l} \ll \min_{r \in S} \frac{1}{\tau_r}.
$$
To model RCP without the effects of delays, we
use~(\ref{eqn:RCPx}-\ref{eqn:smallQ}), setting $\tau_r = 0$ for all
$r \in S$, but leaving all other parameters fixed.

Ideally we would like to show that our delay-free model of RCP, at
equilibrium, is globally stable. Unfortunately, as the following
example demonstrates, we cannot expect the equilibrium points
of~(\ref{eqn:RCPx}-\ref{eqn:smallQ}) to be unique, or even isolated.
Consider a network consisting of two identical links, $l, j$ and one
route $r = \{l, j\}$. Then, since $p_l(\cdot)$ is strictly
increasing, there exists a unique $y$ such that
$$\frac{\alpha_l}{d_l C_l} ( C_l - y ) - \frac{\beta_l p_l(y)}{d_l^2 C_l} = 0.$$
By inspection, there are equilibrium points at $R = (y, y')$ and $R
= (y', y)$ for all $y' \geq y$.

Alternatively, consider the situation where $l$ and $j$ are not
completely identical, and $C_j > C_l$. Then, there exists a unique
$y'$ such that
$$\frac{\alpha_j}{d_j C_j} ( C_j - y' ) - \frac{\beta_j p_j(y')}{d_j^2 C_j} = 0.$$
Now, $y' > y$, so, if $x_r \leq y$ then $\dot{R}_j$ will be strictly
positive. However, if $x_r > y$ then $\dot{R}_l$ will be strictly
negative. Thus, for this example, no equilibrium point exists.

So, there may be an entire manifold of equilibrium vectors for $R$,
or there may be no equilibrium point at all. However, the same is
not true for the equilibrium vector of flow rates $x$.

\begin{theorem}
Suppose that $x(t)$ evolves according
to~(\ref{eqn:RCPx}-\ref{eqn:smallQ}), with $\tau_r = 0$ for all $r
\in S$. Then, there exists a unique vector $\ovx$ such that $x(t)
\rightarrow \ovx$ as $t \rightarrow \infty$.
\end{theorem}
\begin{proof}

We prove this result recursively for the more general system where,
for all $l \in J$, we replace~(\ref{eqn:RCPr}) with
\beqa \dot{R}_l(t) = R_l(t) \left( u_l(t) + \frac{\alpha_l}{d_l C_l}
( C_l - y_l(t) ) - \frac{\beta_l p_l(y_l(t))}{d_l^2 C_l}
\right)^+_{R_l(t)}, \label{eqn:RCPruglob} \eeqa
where $u_l(t) \rightarrow 0$ as $t \rightarrow \infty$.

Let $N_l$ be the number of $r \in S$ such that $l \in r$. We set
$\ovy_l$ equal to the unique value such that $\dot{R}_l(t) = 0$ when
$y_l(t) = \ovy_l$, and set $\ovR = \min_{l \in J} \ovy_l / N_l$. We
shall show that, for all $r \in S$ with $\ovR = \ovy_l / N_l$ for
some $l \in r$, $x_r \rightarrow \ovR$ as $t \rightarrow \infty$.
This allows us to remove such an $x_r(t)$ from the system, by
replacing it with $\ovR$ plus a vanishing term which we incorporate
into the $u_l(t)$, for each $l \in r$. Since this will always remove
at least one $r$ from $S$, this is sufficient to prove our result.

It remains to show that $x_r \rightarrow \ovR$ for $r \in S$ with
$\ovR = \ovy_l / N_l$, for some $l \in r$. Now, for any $\epsilon >
0$ there exists a $T$ such that
$$
|u_l(t)| < \epsilon \frac{\alpha_l}{d_l C_l},
$$
for all $l \in J$, $t > T$. From~(\ref{eqn:RCPruglob}) we see that,
if, for $t > T$, for any $l \in J$, $R_l(t) < (\ovy_l / N_l) - 2
\epsilon$, then $\dot{R}_l(t) > \delta$ for some $\delta > 0$. Thus,
for some $T'$, for all $t > T'$, $l \in J$, $R_l(t) \geq (\ovy_l /
N_l) - 2 \epsilon$.

Let us assume that $\epsilon$ is small enough that, for any $j \in
J$ with $(\ovy_j/N_j) > \ovR$,
$$\frac{\ovy_j}{N_j} - \ovR > \left(2 + 2 \max_{l \in J} N_l \right) \epsilon.$$

Now, let $r \in S$ be such that $\ovR = \ovy_l/N_l$ for some $l \in
r$. Let $r'$ be the set of $l \in r$ with $\ovR = \ovy_l / N_l$.
Suppose that, for some $t > T'$,
$$
\min_{l \in r'} R_l(t) > \ovR + 2 \epsilon \max_{j \in J} N_j.
$$
If $x_r(t) = R_j(t)$ and $j \in r'$ then $x_r(t) > \ovR + 2 \epsilon
\max_{j \in J} N_j$. Otherwise $x_r(t) = R_j(t)$ with $j \not\in
r'$, but since $t > T'$, $x_r > \ovR + 2 \epsilon \max_{j \in J}
N_j$ still holds. For any $l \in r'$, $x_r(t)$ is sufficiently high
that $y_l(t)$ must be greater than $N_l \ovR + 2 \epsilon$. Thus,
$\dot{R}_l(t) < - \delta$ for some $\delta > 0$. Therefore, for some
$T''$, for all $t > T''$,
$$\min_{l \in r'} R_l(t) \leq \ovR + 2 \epsilon \max_{j \in J} N_j.$$
Hence, for all $t > T''$,
$$\left| x_r(t) - \ovR \right| \leq 2 \epsilon \max_{j \in J} N_j.$$

Since $\epsilon$ was arbitrary, the result follows.
\end{proof}

Note, the vector $\ovx$ will be close to that of the max-min fair
allocation of flow rates, because of its construction. However, the
presence of the function $p_l(\cdot)$ in~(\ref{eqn:RCPr}) means that
link capacity will not be fully utilised at equilibrium.

\subsection{Local stability with propagation delays}

In this subsection, we derive conditions for the local stability
of~(\ref{eqn:RCPx}-\ref{eqn:smallQ}) when propagation delays are
present. Our result relies upon the assumption that there is only
one bottleneck link along each route, that is for each $r \in S$
there is only one $l \in r$ such that $x_r(t) = R_l(t)$ at
equilibrium.

For each $r \in S$, $l \in J$, we let $\ovx_r$ be the equilibrium
value of $x_r(t)$ and $\ovR_l$ be the maximum of $\ovx_s$ for all $s
\in S$ such that $l \in s$. For each $l \in J$, we let $\ovy_l$ be
the unique value such that $\dot{R}_l(t) = 0$ whenever $y_l(t) =
\ovy_l$. We can assume, without loss of generality, that for all $l
\in J$,
$$\ovy_l = \sum_{r : l \in r} \ovx_r.$$
Otherwise, assuming the system is always local to equilibrium,
$R_l(t)$ will simply continually increase. If $R_l(t)$ is initially
large enough then $\max_{r \in S} x_r(t) < R_l(t)$ for all time and
thus, $R_l(t)$ has no effect on the rest of the system, and can be
ignored.

Now, we have assumed that, for all $r \in S$, $l \in r$, if $\ovx_r
= \ovR_l$, then $\ovR_j > \ovR_l$ for all $j \in r$, $j \neq l$.
Furthermore, by definition, if $\ovx_r \neq \ovR_l$, we must have
$\ovx_r < \ovR_l$. Thus, for all $r \in S$, $l \in r$, whenever the
system is close to equilibrium, either $x_r(t) = R_l(t -
\tau_{lr})$, or else $\ovx_r < \ovR_l$ and $R_l(t - \tau_{lr})$ has
no effect on $x_r(t)$. So, we can isolate each $l \in J$ and model
the effect of the rest of the system on $R_l(t)$ as a vanishing
perturbation. This allows us to find conditions for stability for
the overall system using a recursive argument.

Accordingly, we initially restrict our attention to the single link
case, $J= \{l\}$, and we consider the following generalisation
of~(\ref{eqn:RCPr}),
\be \dot{R}_l(t) = -R_l(t) \bigg( f_l \Big( y_l(t) - N_l \ovR_l
\Big) + u_l(t) \bigg)^+_{R_l(t)}, \label{eqn:locRCP} \ee
where $y_l(t)$ is defined as in~(\ref{eqn:flow}), $f_l(\cdot)$ is an
increasing differentiable function with $f_l(0) = 0$ and $u_l(t)
\rightarrow 0$ as $t \rightarrow \infty$. When the recursive
argument is complete, the vector $u(t)$ represents the behaviour of
$R_j(t)$ for $j \in J$ such that $\ovR_j < \ovR_l$.

\begin{theorem}
\label{th:RCPtech1} Consider the case where the network consists of
only a single link, $J = {l}$, with flow rates
following~(\ref{eqn:locRCP}). If
$$f_l'(0) \ovR_l \sum_{r \in S} \tau_r < 1,$$
then for all $\epsilon$, there exists $a$ and $u$ such that, if
$\left| R_l(t) - \ovR_l \right| \leq a$ for all $t \leq 0$ and
$\left| u_l(t) \right| < u$ for all $t$, then $\left| R_l(t) -
\ovR_l \right| \leq \epsilon$ for all $t$, and $R_l(t) \rightarrow
\ovR_l$ as $t \rightarrow \infty$.
\end{theorem}

\begin{proof}

Let $\tau_l$ be $\max_{r \in S} \tau_r$, the maximum round trip
time.

Suppose for some $t$, $\gamma \leq 1$, $a'$ and $u'$; $\left| R_l(t)
- \ovR_l \right| = \gamma a'$, $\left| R_l(t') - \ovR_l \right| \leq
a'$ and $\left| u_l(t') \right| < u' a'$ for all $t' \in \left[ t -
2 \tau_l, t \right]$. Then, for all $t' \in \left[ t - \tau_l, t
\right]$,
\beqas \left| \dot{R}_l(t') \right| \leq \left( \ovR_l + a' \right)
\Big( u' a' + f_l' (0) |S| a' + o(a') \Big). \eeqas
Thus, for all $r \in S$,
\be \left| R_l(t - \tau_r) - R_l(t) \right| \leq \tau_r \ovR_l f_l'
(0) |S| a' + \tau_r \left( \ovR_l + a' \right) a' u'  + o(a'). \ee
So,
\begin{eqnarray*}
\ds\frac{\dot{R}_l(t)}{R_l(t)} \hspace{-0.2cm} & = & \hspace{-0.2cm}
f_l'(0) \left( |S| \ovR_l - y_l(t) \right) - u_l(t) + o(a')\\
& = & \hspace{-0.2cm} f_l'(0) |S| \left( \ovR_l - R_l(t) \right) -
u_l(t) + f_l'(0) \sum_{r \in S} \Big( R_l(t) - R_l(t - \tau_r) \Big)
+ o(a')\\
& = & \hspace{-0.2cm} f_l'(0) |S|  \gamma a' + \delta,
\end{eqnarray*}
where
\begin{eqnarray*}
\left| \delta \right| \hspace{-0.2cm} & \leq &  \hspace{-0.2cm}
f_l'(0) \ovR_l \sum_{r \in S} \tau_r f_l' (0) |S| a' + |S| \tau_l
\left( \ovR_l + a' \right) a' u' + o(a')
\\
& = & \hspace{-0.2cm} f_l' (0)^2 |S|^2 \ovR_l d_l a' + |S| \tau_l
\left( \ovR_l + a' \right) a' u' + o(a').
\end{eqnarray*}
Since $f_l' (0) |S| \ovR_l d_l < 1$, there exists a value of $a <
\epsilon$, $u'$ and $\gamma < 1$ such that, if $a' \leq a$, then $|
\delta |$ is guaranteed to be less than $\gamma f_l'(0) |S| a'$. In
which case, $\dot{R}_l(t)$ must have the same sign as $\ovR_l -
R_l(t)$.

So, we can take $u = u' a$, with $a$ as given above, and if $\left|
R_l(t) - \ovR_l \right| \leq a$ for all $t \leq 0$ and $\left|
u_l(t) \right| < u$ for all $t$, then $\left| R_l(t) - \ovR_l
\right| \leq \epsilon$ for all $t$.

Furthermore, for all $a' < a$, there exists a time $T$ such that,
for all $t > T$, $|u_l(t)| < u' a'$. From the above analysis, we
know that after $T$, if $R_l(t)$ is beyond $a'$ of $\ovR_l$, then it
will converge to $\ovR_l$ at a rate of $\gamma$. Thus, there is some
$T' > T$ such that, $\left| R_l(t) - \ovR_l \right| \leq a'$ for all
$t > T'$. Since $a'$ was arbitrary, $R_l(t) \rightarrow \ovR_l$ as
required.
\end{proof}

We now consider the general network case.

\begin{theorem}
\label{th:RCPlocnet} Under~(\ref{eqn:RCPx}-\ref{eqn:smallQ}), the
equilibrium point $\ovR$ is locally asymptotically stable provided
that, for each $l \in J$,
\be \Bigl( \frac{\alpha_l}{d_l C_l}  +  \frac{\beta_l
p'_l(\ovy_l)}{d_l^2 C_l} \Bigr) \ \ovR_l \sum_{r : l \in r, \ovR_l =
\ovx_r} \hspace{-0.5cm} \tau_r < 1. \label{eqn:RCPnetdscond} \ee
\end{theorem}
\begin{proof}

Let $\delta$ be the minimum of $\left| \ovR_l - \ovR_j \right| /2$
for $l \neq j \in J$. Now, if, for all $t$, all $l \in J$,
$\left|R_l(t) - \ovR_l \right| < \delta$, then each $l \in J$ will
evolve according to
\be \label{eqn:RCPlocgen} \dot{R}_l(t) = -R_l(t) \Big( f_l \big(
y^{eq}_l(t) - \ovy^{eq}_l \big) + u_l(t) \Big), \ee where $u_l(t)$
represents the effect of $R_j(t)$ for all $\ovR_j < \ovR_l$, and
$$
y^{eq}_l(t) = \hspace{-0.2cm} \sum_{r : l \in r, \ovx_r = \ovR_l}
\hspace{-0.5cm} x_r(t - \tau_{rl}),
$$
with $\ovy^{eq}_l$ equal to the equilibrium value of $y^{eq}_l(t)$.
For each $l \in J$, we have
$$f_l(w) =
\frac{\alpha_l}{d_l C_l} ( w + \ovy_l - C_l) + \frac{\beta_l
p_l(\ovy_l + w)}{d_l^2 C_l}.
$$

By definition,~(\ref{eqn:RCPlocgen}) is an example
of~(\ref{eqn:locRCP}), for the reduced network formed by $l$ and all
routes $r \in S$ with $l \in r$ and $\ovx_r = \ovR_l$.
Furthermore,~(\ref{eqn:RCPnetdscond}) is precisely the condition of
Theorem~\ref{th:RCPtech1} for this reduced system.

Suppose $\left|R_l(t) - \ovR_l \right| < \delta$ for all $l \in J$,
for all $t$. By differentiability of $p_l(\cdot)$, for any $u$, we
can find $\epsilon$ such that if $\left|R_j(t) - \ovR_j \right| <
\epsilon$ for all $t$, for all $j$ such that $\ovR_j < \ovR_l$, then
$\left| u_l(t) \right| < u$ for all $t$. This allows us to prove
local stability recursively.

We begin with the $l \in J$ such that $\ovR_l$ is maximal. We apply
Theorem~\ref{th:RCPtech1} to find conditions under which $R_l(t)$
converges to $\ovR_l$ and $\left| R_l(t) - \ovR_l \right| < \delta$
for all $t$. This gives us $a$ and $u$, where the initial conditions
of $R_l(t)$ should be within $a$ of $\ovR_l$, and $u_l(t)$ should
tend to zero and always be bounded by $u$. We can find an $\epsilon
< \delta$ so that if $\left| R_j(t) - \ovR_j \right| < \epsilon$ for
all $t$ and $R_j(t)$ converges to $\ovR_j$ for all $j \neq l$, then
$\left| u_l(t) \right| < u$ for all $t$ and $u_l(t)$ converges to
$0$. So, we set $R_l(t) \in \left[ \ovR_l - a, \ovR_l + a \right]$
for $t \leq 0$ as our initial condition for $l$, remove $l$ from the
network and repeat this process for $\delta' = \epsilon$. Since we
remove one link each time, eventually we will find suitable initial
conditions for the entire network.
\end{proof}

Note, each link does not necessarily need to keep track of which
flows are under its control in order to
meet~(\ref{eqn:RCPnetdscond}). The condition holds if, for all $l
\in J$,
$$
\left( \frac{\alpha_l}{d_l }  +  \frac{\beta_l p'_l(\ovy_l)}{d_l^2 }
\right) \frac{\ovy_l}{C_l} d_l^{p} < 1,
$$
where $d_l^p$ is the average per \emph{packet} round trip time,
$$
d_l^p = \frac{1}{\ovy_l} \sum_{r : l \in r} \ovx_r \tau_r.
$$
This suggests that it should be $d_l^p$ not $d_l$ that appears in
the RCP controller~(\ref{eqn:RCPr}). If, for each $l \in J$ we let
$\gamma_l$ be such that $p'_l(\ovy_l) = \gamma_l p_l(\ovy_l) /
\ovy_l$ then,
$$
p'_l(\ovy_l) = \gamma_l \frac{\alpha_l d_l}{\beta_l} \frac{C_l -
\ovy_l}{\ovy_l} < \frac{\gamma_l \alpha_l d_l}{\beta_l},
$$
thus,~(\ref{eqn:RCPnetdscond}) is ensured if, for all $l \in J$, \be
\label{eqn:RCP:nicenetcond} \alpha_l = \frac{d_l^p}{d_l (1 +
\gamma_l)}. \ee
This parameter choice scheme is attractive, because it is fairly
decentralised and only requires local information. Each link needs
only to measure, estimate or be informed of the round trip times for
packets passing through that link. However, these results rely on
the weak assumption that each route has only \emph{one} bottleneck
link. However, any network with multiple bottleneck routes can
easily be transformed into one without, for example, by making a
small change in the parameters $\beta_l$, $l \in J$. Preliminary
numerical results suggest that Theorem \ref{th:RCPlocnet} may hold
in general; but finding an analytic result to confirm this, or a
counter-example to disprove it, remains an open problem.

In the next section we investigate the impact of loss of local
stability in RCP.

\section{Local bifurcation analysis}

A key focus in the choice of parameters for any congestion control
proposal is to ensure that they lead to a stable equilibrium. Most
proposals for congestion control, for example see \cite{srikant},
lead to the analysis of non-linear time delayed dynamical systems.
For such non-linear systems, typically sufficient conditions for
local stability guide parameter choices. Following local stability,
a local bifurcation theoretic analysis can make us comfortable in
running the system close to the \emph{edge} of stability.

We first recapitulate a result about the loss of local stability in
a non-linear retarded functional differential equation and then use
it to analyse the RCP fluid model. Following the analysis in
\cite{raina2005} we outline a local Hopf bifurcation result for the
following non-linear delay equation
\beqa \dot{u}(t) = \eta \big( -\xi_{x} u(t-\tau) \pm \xi_{xy} u(t)
u(t-\tau) \big), \label{a1} \eeqa
where $\eta, \tau, \xi_x, \xi_{xy} > 0.$ The parameter $\eta$ has
been intentionally introduced to just tip the above equation over
the edge of (local) stability. This exogenous, non-dimensional,
parameter will \emph{act} as our bifurcation parameter.

We may state the following about equation (\ref{a1}).

\begin{theorem} A necessary and sufficient condition for local
stability is
$$
\eta \xi_{x} \tau < \pi/2,
$$
and treating $\eta$ as the bifurcation parameter, the first local
Hopf bifurcation occurs with period $4 \tau$ at $\eta = \eta_{c}$,
where $\eta_{c} \xi_{x} \tau = \pi/2.$ Further, as the Hopf
condition is just violated, the equation will always undergo a
super-critical Hopf bifurcation where the amplitude of the stable
bifurcating solutions will be proportional to
$$
\ds\frac{\xi_{x}}{\xi_{xy}} \sqrt{\frac{20 \pi (\eta - \eta_{c})}{ 3
\pi - 2 }}.
$$
\label{hopf}
\end{theorem}

\subsection{RCP: single link, single delay model}

In our bifurcation analysis, for the sake of simplicity, we shall
leave the queuing term out of the model by taking $\beta = 0$.
Consider the following single link, single delay, RCP fluid model
\beqa \dot{R}(t) =  \eta R(t) \left( \frac{\alpha}{C \tau} \left(C -
y(t) \right) \right), \label{rcp1} \eeqa
where $y(t)= \sum_{s} R(t-\tau)$ is the aggregate load at the link
and $\eta$ is the non-dimensional bifurcation parameter. Define
$u(t) = R(t) - \ovR $, and take a Taylor expansion of (\ref{rcp1})
to obtain
\beqas \dot{u}(t) {\hspace{-0.5cm}}&&= - \ \eta \frac{\alpha}{\tau}
u(t-\tau) - \eta \frac{\alpha}{ \tau \ovR} u(t) u(t - \tau). \eeqas
Now using Theorem \ref{hopf}, we may state the following about
equation (\ref{rcp1}). A necessary and sufficient condition for
local stability is
$$
\eta \alpha < \pi/2,
$$
and treating $\eta$ as the bifurcation parameter, the first local
Hopf bifurcation occurs with period $4 \tau$ at $\eta = \eta_{c}$,
where $\eta_{c} \alpha = \pi/2.$ If the Hopf condition is just
violated, the equation will \emph{always} undergo a super-critical
Hopf bifurcation where the amplitude of the stable bifurcating
solutions will be proportional to
$$
\ovR \sqrt{\frac{20 \pi (\eta - \eta_{c})}{ 3 \pi - 2 }}.
$$
We highlight that equation (\ref{rcp1}) cannot produce a
sub-critical Hopf bifurcation. However, in the bifurcation analysis
we have omitted any additional non-linear effects that would arise
if $\beta > 0$ in the RCP model.

We now summarise our contribution in this paper, and outline some
avenues for further research.

\section{Conclusion}

It is noteworthy to observe that the small buffer regime has allowed
us to tackle the question of stability for a large network with an
arbitrary topology. First, we proved that the RCP fluid model is
globally stable in the absence of propagation delays. Then, we
derive a set of conditions for local stability when arbitrary
heterogeneous propagation delays are present. The network delay
stability result relies upon the weak assumption that, at
equilibrium, there is only one bottleneck link along each route. An
interesting avenue for research would be to show that either this
result may hold in general, or provide a counter-example to disprove
it. Finally in a single link, single delay, model we investigate the
impact of the loss of local stability in a special case (setting
$\beta = 0$) where we show that the RCP algorithm would
\emph{always} give rise to an innocuous looking super-critical Hopf
bifurcation.

As RCP aims for max-min fairness it is appropriate to first consider
a model that embodies the original formulation, as we did in this
paper. We now outline some natural avenues for further research.

\emph{Fairness and Stability.} A conveniently parameterized family
of $\alpha$-fair rate allocations was introduced in~\cite{mw}. The
parameter $\alpha$ lies in the range $(0, \infty)$, and the cases
$\alpha \rightarrow 0$, $\alpha = 1$ and $\alpha \rightarrow \infty$
correspond respectively to an allocation which achieves maximum
throughput, is proportionally fair or is max-min fair~\cite{mw}. In
this paper we have only considered a max-min fair allocation
mechanism, as has been originally specified \cite{rcp}. An immediate
direction of further research would be to incorporate different
notions of fairness into the RCP framework, and analyse the
stability of such networks.

\emph{Impact of the RCP parameters: $\alpha_{l}$ and $\beta_{l}$.}
We have only considered the limiting case of very small buffers. The
choice of $\beta$ impacts the rate at which the queue is drained.
Small values of $\beta$ drain the queue slowly; so with small
$\beta$ and for large enough buffers it is appropriate to model the
queue as a saturated integrator. However, for large values of
$\beta$, the queue may drain fast enough so that at the timescale of
operation of the congestion control protocols, finer queueing
theoretic models may have to be developed. Such queuing models would
certainly contribute to the non-linearity in the RCP dynamical
system, and hence impact the stability of the time delayed network.
Further, the results from local bifurcation theory could also be
subtle. In our bifurcation theoretic analysis we set $\beta = 0$,
thus removing any non-linearity that may arise from the queue.
\\

\noindent {\bf Acknowledgements} \\
The authors are grateful to Frank Kelly for comments on earlier
drafts and acknowledge funding provided by the EPSRC grant
GR/S$86266/01$. The usual caveat applies.

\vfill

\end{document}